\documentclass[aps,prl,twocolumn,showpacs,superscriptaddress]{revtex4-1}
\usepackage[latin1]{inputenc}
\usepackage{graphicx}
\usepackage{epstopdf}
\usepackage{epsfig}
\usepackage{xfrac}
\usepackage{indentfirst}
\usepackage{amsmath, amsthm, amssymb} 
\usepackage{bm}
\input epsf
\usepackage{float}
\usepackage[colorlinks=true, a4paper=true, pdfstartview=FitV, linkcolor=blue, citecolor=blue, urlcolor=blue]{hyperref}
\usepackage{braket}
\begin{document}  
\title{Turbulence Excitation in Counter-Streaming Paraxial Superfluids of Light} 
\author{Jo\~{a}o D. Rodrigues}
\email[Corresponding author:]{j.marques-rodrigues@imperial.ac.uk}
\affiliation{Physics Department, Blackett Laboratory, Imperial College London, Prince Consort Road, SW7 2AZ, United Kingdom}
\affiliation{Instituto de Plasmas e Fus\~{a}o Nuclear, Instituto Superior T\'{e}cnico, Universidade de Lisboa, 1049-001 Lisbon, Portugal}
\author{Jos\'{e} T. Mendon\c ca}
\affiliation{Instituto de Plasmas e Fus\~{a}o Nuclear, Instituto Superior T\'{e}cnico, Universidade de Lisboa, 1049-001 Lisbon, Portugal}
\author{Hugo Ter\c cas}
\affiliation{Instituto de Plasmas e Fus\~{a}o Nuclear, Instituto Superior T\'{e}cnico, Universidade de Lisboa, 1049-001 Lisbon, Portugal}
\begin{abstract}
\begin{footnotesize}
Turbulence in the quantum (superfluid) regime, similarly to its classical counterpart, continues to attract a great deal of scientific inquiry, due to the yet high number of unresolved problems. While turbulent states can be routinely created in degenerate atomic gases, there is no generic scheme to produce turbulence in fluids of light. Under paraxial propagation, light in bulk nonlinear media behaves as a two-dimensional superfluid, described by a nonlinear Schr\"{o}dinger equation formally equivalent to the Gross-Pitaevskii model of a weakly interacting Bose gas, where photon-photon interactions are mediated by a third order (Kerr) nonlinearity. Here, we develop the theory describing the onset of a kinetic instability when two paraxial optical fluids with different streaming velocities interact via the optical nonlinearity. From numerical simulations of the nonlinear Schr\"{o}dinger equation, we further characterize the onset of the instability and describe its saturation in the form of vortex nucleation and excitation of turbulence. The experimental observation of such effects is also discussed. The class of instabilities described here thus provide a natural route towards the investigation of quantum (superfluid) turbulence, structure formation and out-of-equilibrium dynamics in superfluids of light.
\end{footnotesize}
\end{abstract}
\maketitle
\section{Introduction}
\par 
Turbulence is an ubiquitous phenomenon of nature, characterized by a nontrivial and rather chaotic cascading of energy between different scales~\cite{Tennekes, Pope}. A great deal of our current understanding of turbulence can be traced back to the pioneer work of Richardson introducing the concept of the energy cascade and the subsequent quantitative framework developed by Kolmogorov~\cite{Frisch}. Nevertheless, a microscopical understanding of the mechanisms leading to turbulent flow are yet unknown, even in the classical regime~\cite{Falkovich2006}. On the other hand, quantum gases emerge as an exciting new paradigm for the investigation of (quantum) turbulence~\cite{Yepez2009, Navon2016}, where the superfluid~\cite{Onofrio2000, Desbuquois2012} behavior imposes the quantization of the total vorticity content.
\par 
Quantum gases owe most of their peculiar nature to the wave-like character of matter emerging at low temperatures, when the de Broglie wavelength becomes larger than the inter-particle distance. In these conditions, an emergent macroscopic wavefunction describes the entire collection of particles, such as in the case of atomic Bose-Einstein condensates~\cite{Dalfovo1999}. Interestingly, light can also be described as a quantum fluid, like in semiconductor microcavities, where gases of exciton-polaritons undergo Bose-Einstein condensation~\cite{Kasprzak2006, Deng2010}. Here, while spatial confinement turns photons into massive particles, mixing with matter degrees of freedom provides effective photon-photon interactions. With the observation of superfluidity~\cite{Amo2009_1} and the formation of vortices or solitons~\cite{Lagoudakis2008, Amo2011}, these quantum fluids of light~\cite{Carusotto2013} became the subject of great scientific interest.
\par 
Yet in the context of nonlinear optics, a rather different approach to superfluidity arises in bulk nonlinear media under paraxial optical propagation~\cite{Chiao1999, Larre2015_1, Vocke2015}. The (two-dimensional) transverse dynamics of light is described by a nonlinear Schr\"{o}dinger (NLS) equation similar to the Gross-Pitaevskii equation for the macroscopic wavefunction of a weakly interacting Bose gas~\cite{Dalfovo1999}. Here, however, the time evolution is formally mapped onto the longitudinal propagation direction, corresponding to a conservative (Hamiltonian) dynamics in this transformed coordinate system. Besides the cancellation of the drag force onto an obstacle~\cite{Michel2018}, the hallmark of superfluid behavior, effects such as the formation of vortices~\cite{Swartzlander1992}, Rayleigh-Taylor instabilities~\cite{Shu2012} and kinetic Bose-Einstein condensation~\cite{Sun2012} have been observed under this paraxial geometry.
\par 
In this work, we describe the onset of a two-stream instability when two counter-flowing paraxial fluids interact via the optical nonlinearity. We begin by introducing the hydrodynamic model of the two-dimensional optical fluid and analyze its dispersion and stability. Under specific conditions, an instability can be triggered, rooted in the resonant energy transfer from the streaming flow to the elementary (Bogoliubov) excitations. The onset of the unstable regime is described and numerical simulations of the nonlinear Schr\"{o}dinger equation allow us to characterize its nonlinear stages. We show that quantum turbulence can be excited as a result of the counter-streaming flow.
\par 
Optical superfluid systems such as those described here are optimal platforms to investigate quantum turbulence. As it is clear now, 2D quantum turbulence~\cite{Bradley2012, Reeves2013, Billam2014} is fundamentally different than its 3D counterpart. Paraxial superfluids are the only available systems which are genuinely 2D, whereas in typical atomic BECs the 2D character is only approximate. Also, experiments on paraxial optical superfluids~\cite{Michel2018, Fontaine2018} are extremelly more simple than those required to produce atomic BECs, while maintaining a high level of control. Finally, it has been demonstrated that one can generate optical fluids of an extent much larger than the healing lenght~\cite{Santic2018}, thus, in principle, allowing for a high number of vortices to be generated while granting access to the dynamics across a large range of scales. Here, we go one step further by introducing and describing a simple way to generate quantum vortices and turbulence in these paraxial fluids of light.
\section{Optical Hydrodynamics and Two-Stream Instability}
\label{sec:theory}
\par 
Under paraxial propagation, the electric field of a monochromatic optical pulse $E(\mathbf{r}, t) = \psi (\mathbf{r}_{\perp}, z) e^{ink_0 z - i \omega_0 t}$, with $\mathbf{r}_{\perp} = (x,y)$, $k_0 = \omega_0 / c$ the (vacuum) wavenumber of the carrier wave, $\omega_0$ its frequency and $c$ the speed of light, obeys the nonlinear Schr\"{o}dinger (NLS) equation~\cite{Carusotto2013, Vocke2015}
\begin{equation}\label{eq:nlse_1}
i \frac{\partial \psi}{\partial z} = - \frac{1}{2nk_0} \nabla_{\perp}^2 \psi - \frac{k_0}{2n} \chi^{(3)} \lvert \psi \rvert^2 \psi,
\end{equation}
with $\psi (\mathbf{r}_{\perp}, z)$ being the slowly varying field amplitude. The third order (Kerr) susceptibility $\chi^{(3)}$ quantifies the optical nonlinearity, with $n$ the linear index of refraction. The paraxial approximation is rooted in the idea of the spatial variations of the field envelope being much slower than those of the carrier wave, such that the transverse dynamics is highly suppressed~\cite{Carusotto2013}. The optical nonlinearity is at the origin of contact interactions of strength $g= - \frac{k_0}{2n} \chi^{(3)}$ or, equivalently, the photon-photon potential $V(\mathbf{r}_{\perp} , z) = g \lvert \psi (\mathbf{r}_{\perp} , z) \rvert^2$, with $\chi^{(3)} > 0$ ($\chi^{(3)} < 0$) corresponding to attractive (repulsive) interactions.
\par 
The NLS equation can be mapped onto an hydrodynamic formulation by prescribing $\psi = \sqrt{\rho}e^{i \phi}$, with the fluid density (light intensity) and velocity field defined as $\rho = \lvert \psi \rvert^2 = \lvert E \rvert^2$ and $\mathbf{v} = (nk_0)^{-1}\mathbf{\nabla}_{\perp} \phi$, respectively. These satisfy the hydrodynamic equations
\begin{equation}\label{eq:hydro1}
\frac{\partial \rho}{\partial z} + \mathbf{\nabla}_{\perp} \left( \rho \mathbf{v} \right) = 0, \qquad \text{and}
\end{equation}
\begin{equation}\label{eq:hydro2}
\frac{\partial \mathbf{v}}{\partial z} + \left( \mathbf{v} \cdot \mathbf{\nabla}_\perp \right) \mathbf{v} = -\frac{g}{nk_0} \mathbf{\nabla}_\perp \rho + \frac{1}{2} \mathbf{\nabla}_\perp \left( \frac{1}{\sqrt{\rho}} \nabla^2_\perp \sqrt{\rho} \right),
\end{equation}
where the coordinates have been re-scaled as $(x, y, z) \rightarrow nk_0 (x, y, z)$. The wave character of the system manifests itself through the last term in Eq.~(\ref{eq:hydro2}), usually referred to as the Bohm potential (or quantum pressure). Note that, due to the $t \leftrightarrow z$ mapping, flow speeds have no units and correspond to angles of light rays, quantifying transverse to longitudinal propagation.
\begin{figure}
\centering
\includegraphics[scale=0.76]{fig1}
\caption{(color online) Stability diagram in the $(q \xi, \beta)$ plane. The upper subplane ($\beta > 2$) defines the region of fast flows, while the lower one ($\beta < 2$) depicts the slow streaming regime, where a dynamical instability is expected at all wavenumbers below $\beta$.}
\label{fig:unstable_region}
\end{figure}
\par 
We now consider the case of two optical fluids where, without lost of generality, one is considered to be at rest, while the other is flowing at a velocity $\mathbf{v}_0$. We note that, while in atomic BEC literature, a two-fluid system usually refers to a mixture of different spin components~\cite{Hall1998, Lin2011}, here the two fluids are of the same type. The hydrodynamic quantities can be expanded as $\rho_{\text{1,2}} = \rho_0/2 + \delta \rho_{\text{1,2}}$, $\mathbf{v}_1 = \delta\mathbf{ v}_1$, $\mathbf{v}_2 = \mathbf{v}_0 + \delta\mathbf{ v}_2$ and the corresponding governing equations linearized by retaining only first order terms in the perturbations. As demonstrated in Ref.~\cite{Abad2015}, the hydrodynamic description of a two-component superfluid in terms of the velocity fields $\mathbf{v}_1$ and  $\mathbf{v}_2$, is equivalent to the NLS equation for the superposition $\psi = \psi_1 + \psi_2$. This equivalence will be useful for the numerical simulation of the two-stream instability, described in the next section. 
\par
The linear dynamics are evaluated by Fourier decomposition, $\delta \rho_{\text{1,2}} (\mathbf{r}_\perp, z) = A_{\text{1,2}} e^{i \mathbf{q} \cdot \mathbf{r}_\perp - i \Omega z}$, with $A_\text{i}$ the mode amplitude and $\mathbf{q}$ and $\Omega$ the transverse wavevector and frequency of the elementary excitations, respectively, related by the dispersion relation 
\begin{equation}\label{eq:two_fluid_dispersion_relation}
1- \frac{1}{2}c_s^2q^2 \left[ \frac{1}{\Omega^2-q^4/4} + \frac{1}{\left(\Omega - \mathbf{v}_0 \cdot \mathbf{q} \right)^2 - q^4/4} \right] = 0,
\end{equation}
with $c_s = \sqrt{\rho_0 g / nk_0} = n^{-1} \sqrt{- \chi^{(3)} \lvert E_0 \rvert^2/2}$ the two-fluid speed of sound. In the case of a single optical component of density $\rho_0$ and zero drift velocity, $\Omega ^2 = c_s^2 q^2 +  q^4 / 4$, equivalent to the Bogoliubov dispersion relation of a weakly interacting Bose gas, describing an acoustic (fluid) regime at low momenta. This has been recently observed in a paraxial fluid of light~\cite{Fontaine2018}. Due to the $t \leftrightarrow z$ mapping, $\Omega$ has units of inverse length. The linear dispersion is at the origin of superfluidity~\cite{Vocke2015}, related with the suppression of radiation pressure forces~\cite{Larre2015_1}. We can define the healing length $\xi$ as the scale at which the fluid behavior breaks and turns into a single particle (massive photon) regime -- occurring approximately at $q \sim \xi^{-1}$, with $\xi = c_s^{-1}$. In terms of the optical parameters, $\xi = k_0^{-1} \sqrt{-2/ (\chi^{(3)} \lvert E_0 \rvert^2)}$.
\par 
We now turn to the full dispersion relation in Eq. (\ref{eq:two_fluid_dispersion_relation}), whose solutions given by
\begin{eqnarray}\label{eq:two_fluid_frequency}
\Omega \xi^2 &=& \frac{1}{2} \left(q \xi \right) \Bigg[ \beta  \pm \nonumber \\
&  & \sqrt{ 2 + \beta^2 + \left(q\xi\right)^2 \pm 2 \sqrt{1 + 2\beta^2 + \beta^2 \left(q\xi\right)^2}} \quad \Bigg],
\end{eqnarray}
where the stream parameter is defined as $\beta = v_0 / c_s$. Unstable modes (complex roots) exists in two distinct regimes -- see Fig.~(\ref{fig:unstable_region}). On the one hand, for ``slow" flows such that $\beta < 2$, unstable excitations exist for all wavenumbers satisfying $0 < q \xi < \beta$. In regular units, the latter translates to $0 < q < n k_0 \sin{\theta}$, with $\theta$ the angle of optical propagation. As such, a transverse instability will be verified at all wavenumbers smaller than that of the carrier wave projected onto the transverse subspace. On the other hand, for ``fast" flows ($\beta \geq 2$), the instability is restricted to the spectral region $\sqrt{\beta^2-4} < q\xi < \beta$. While the upper boundary is the same as in the slow regime, a lower instability cut-off is now expected, given by the resonance condition between the optical stream and the Bogoliubov modes, namely $ \beta (q \xi) / 2 = \Omega_B$, with $\Omega_B^2\xi^4 = \left( q \xi \right)^2 + 1/4 \left( q \xi \right)^4$.
\begin{figure}
\centering
\includegraphics[scale=0.76]{fig2}
\caption{(color online) Real (top) and imaginary (bottom) parts of the dispersion relation of the two-fluid system. The full black darker line depicts the unstable mode, while, for the sake of reference, the red lighter and black darker dashed lines represent both the Bogoliubov dispersion and the streaming term $\Omega \xi^2 = \beta (q\xi) / 2$, respectively. From left to right $\beta=1$ (slow), $\beta=2$, $\beta=3$ (fast), with the shaded area marking the unstable region.}
\label{fig:streaming_instability}
\end{figure}
\par 
The dispersion relation in both the slow and fast regimes is depicted in Fig.~(\ref{fig:streaming_instability}). In the stable region, the modes of the resting fluid are described by the Bogoliubov dispersion. The unstable modes, however, acquire the character of the optical stream (mode attraction). For fast streams, the unstable region at high momenta narrows and, in the limit $\beta \rightarrow \infty$ there is only was unstable mode, $q \xi = \beta$. The instability is fully rooted in the acoustic dispersion and fluid-like properties arising from photon-photon interactions. In plasma physics, two-stream instabilities occurs when travelling electrons couple via electrostatic interactions with an ionic background~\cite{Nicholson}. These instabilities, which are often interpreted as inverse Landau damping, are generally described by kinetic equations in both real and momentum space. The latter is equivalently mimicked here with the hydrodynamic equations for a two-component fluid with different streaming velocities. In a more general framework, one can contruct a kinetic description in terms of a Wigner function defined in both real and momentum space which obbeys a Wignar-Moyal equation~\cite{Mendonca2008, Tercas2009}. While this alternative model is completely equivalent to the NLS equation, it describes wave-wave interactions in a more natural way, thus providing an interesting tool for studies on wave turbulence.
\par 
A few notes about other known instability mechanisms in superfluids are in order. In particular, one can consider the quite general class of counter-flow instabilities. In liquid helium, the counter-flow between a superfluid and a normal fluid is known to excite turbulence~\cite{Guo2010, Lushnikov2018}. The presence of the normal component induces velocity shear, such that these instabilities are of the Kelvin-Helmholtz type. The Rayleigh-Taylor constitutes yet another kind of counter-flow, which has been considered in 2D superfluids of light~\cite{Shu2012}. Here, the instability develops at the interface between two fluids of different densities which are accelerated towards each other, thus being fundamentally different from the case investigated here. Counter-flow instabilities between two superfluids have also been investigated in the context of a mixture of Bose-Einstein condensates~\cite{Abad2015, Ishino2011}. Here, however, the instability depends on a miscibility condition between the two components, which relates the intra-- and inter--species interaction strength. Despite the similarities with the two-stream instability, previous results do not transpose directly our system. 
\section{Numerical Simulations}
\label{sec:simulations}
\par 
In order to inquiry on the accuracy of the theoretical model, as well as describing the nonlinear stages of the two-stream instability, we numerically integrate the NLS equation using a split-step Fourier scheme~\cite{Antoine2013}. The grid size is 0.2 $\xi$ with a total number of 1000 grid points, while the integration time step is 0.01 $\xi / c_s$. With a total of 4000 integration points we achieve a simulation time span of 40 $\xi / c_s$, long enough to probe both the instability onset, as well as its nonlinear stages.
\begin{figure*}
\centering
\includegraphics[scale=0.60]{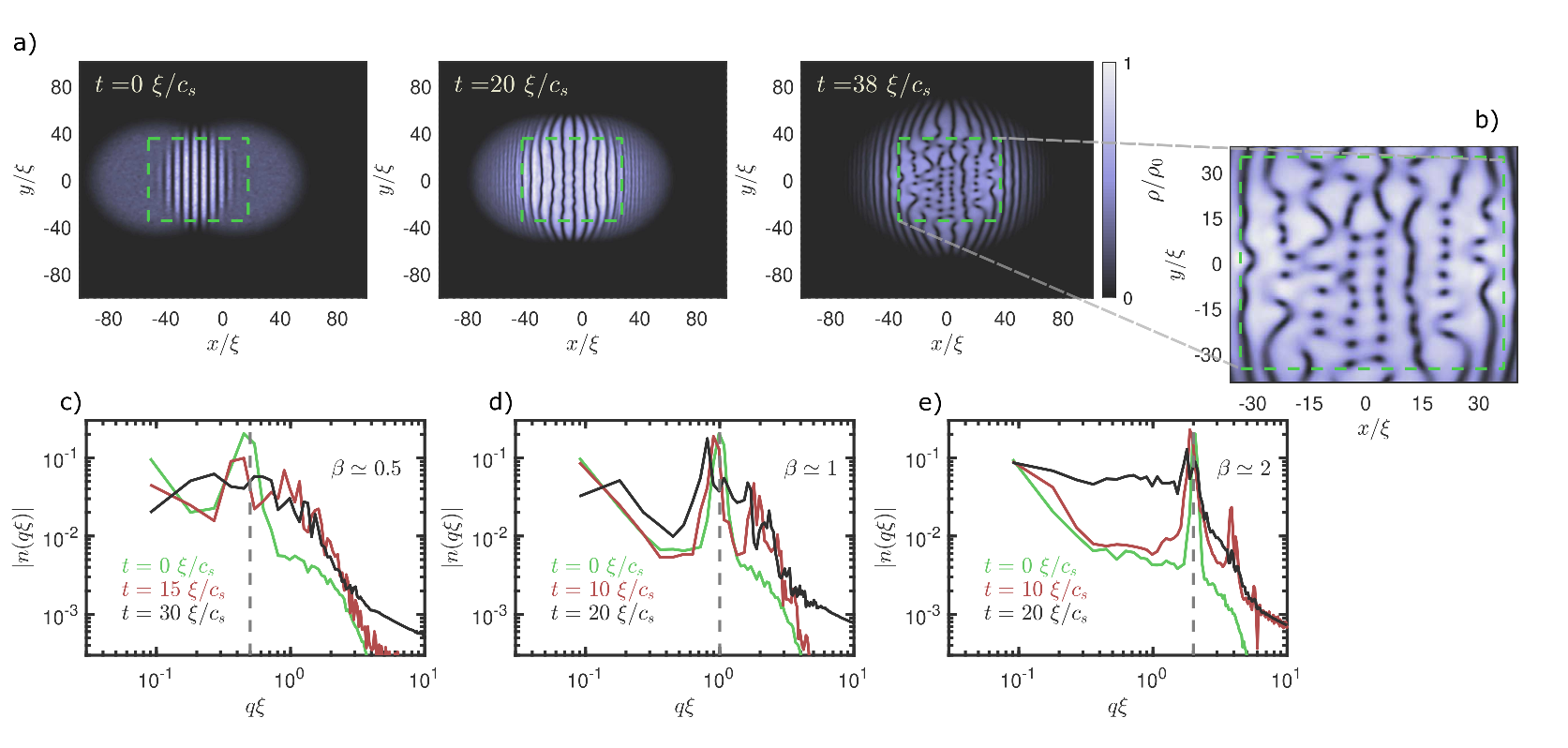}
\caption{(color online) Numerical results on the two-stream instability. Panel a) depicts a typical evolution of the two-fluid density $\rho = \vert \psi \vert^2= \vert E \vert^2$, for $\beta \simeq$ 1. The instability develops slowly as one fluid propagates towards the other. Panel b) shows a zoomed view of the central interaction region at latter stages of the instability, which saturates in the form of vortex-antivortex pairs -- see also Fig.~(\ref{fig:panel2}). The green dashed box enclosures the interaction region where the Fourier spectrum is calculated and depicted in panels c) - e). The dashed vertical lines indicate the wavenumber associated with the initial interference fringes, given by $q\xi = \beta$, which define an energy forcing scale.}
\label{fig:panel}
\end{figure*}
\begin{figure*}
\centering
\includegraphics[scale=0.60]{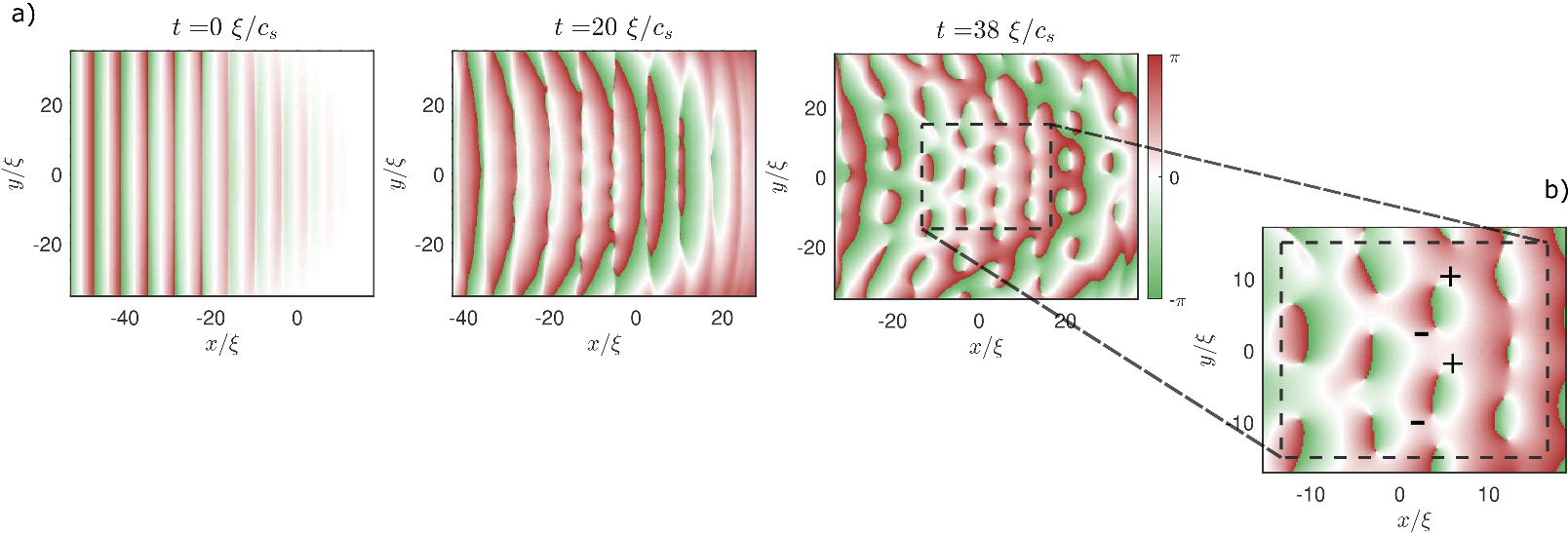}
\caption{(color online) Numerical results on the two-stream instability. Panel a) depicts the phase profile of the two-fluid system, at the same time instants as those depicted in Fig.~(\ref{fig:panel}). Only the region delimited by the green dashed box in Fig.~(\ref{fig:panel}) is plotted here. Panel b) shows a zoomed view of the central interaction region, where one can clearly observe the nucleation of pairs of positive and negative circulation vortices.}
\label{fig:panel2}
\end{figure*}
\par 
To simulate a realistic experimental scenario, we consider two flat-top beams of about 80 $\xi$ in diameter. Each beam is described by the initial wavefunction $\psi_i(t_0) = (1/2) \text{exp} \left[  \left( (x-x_{i0})^2 + (y-y_{i0})^2 \right)^4 / (50)^8  - i \beta_i x \right] N(x, y) $, with $(x_{i0}, y_{i0})$ the initial position of the beams, $i=\lbrace 1,2 \rbrace$, $\left(\beta_1, \beta_2 \right) = \left( \beta, 0 \right)$, and $N(x,y)$ a noisy background computed as a Gaussian process of correlation length $\xi$ and amplitude of 0.05. The single-component NLS Eq.~(\ref{eq:nlse_1}) is integrated starting from the initial condition $\psi(t_0) = \psi_1(t_0) + \psi_2(t_0)$. We implement periodic boundary conditions, although this specific choice does not influence the results, since the beams remain smaller than the full simulation box during the entire simulation timespan. The overlapping region of the two beams is fully equivalent to the two-fluid hydrodynamic model described above, where one component is at rest while the other propagates at a speed $v_0 = \beta c_s$, resulting in the development of a two-stream instability -- see Figs.~(\ref{fig:panel}) and (\ref{fig:panel2}). 
\par 
The homogeneous density of each beam is set to 0.5$\rho_0$ and, upon overlap, the optical fields interfere and fringes develop at $q \xi = \beta$. As shown before, for $\beta <$ 2, the model predicts an instability in the entire infra-red region, corresponding to $q \xi < \beta$. The interference fringes act as a lower cut-off, with the instability always developing at larger scales. This is in agreement with the results depicted in Fig.~(\ref{fig:panel}) -- panels c) to e). At $t=$ 0, the density distribution exhibits a single peak at $q \xi = \beta$ (interference scale) while, for later times, density components at $q\xi < \beta$ grow. This is particular evident for $\beta \simeq$ 2 -- depicted in panel e). The finite size nature of our simulation sets an absolute upper bound on the observable scales, namely the beam size or, in this case, the size of interaction region depicted by the green dashed box in panel a). All the Fourier analysis perform here is conducted only on the density $\rho = \vert \psi \vert^2= \vert E \vert^2$, which is accessible in an experiment as simply the intensity of the light field at the output of the nonlinear material. This allows for a straightforward comparison with the results presented in this work.
\par 
The phase profile depicted in Fig.~(\ref{fig:panel2}) provides further information on the instability mechanism. On the one hand, the interference pattern between the two initial wavepackets resembles solitonic excitations, with longitudinal $2\pi$ phase jumps between fringes. These become unstable and decay into azimuthal $2\pi$ phase jumps, forming vortices in pairs of positive and negative circulation.
\begin{figure}
\centering
\includegraphics[scale=0.74]{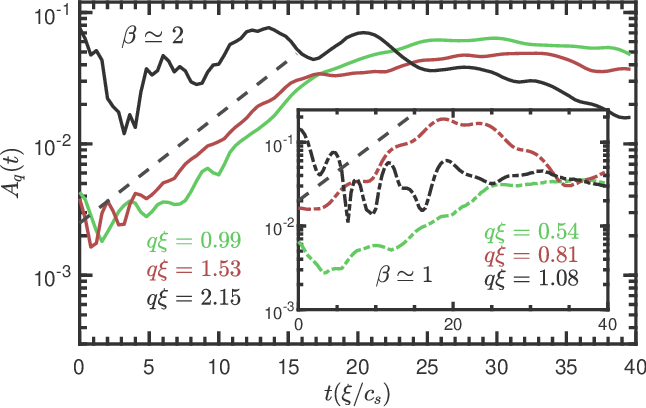}
\caption{(color online) Time-dependent amplitude of stable (black, darker line) and unstable modes (red and green, lighter lines) for two different streaming speeds, $\beta$, shown in the respective panels. The dashed black darker lines indicate the exponentially growing amplitude of the unstable modes. In particular, the growth rates of approximately 0.2 $c_s/\xi$ in the main plot and 0.16 $c_s/\xi$ in the inset are in good agreement with the linear stability analysis -- see Fig.~(\ref{fig:streaming_instability}).}
\label{fig:mode_growth}
\end{figure}
\par 
In order to get further details of the two-stream mechanism, Fig.~(\ref{fig:mode_growth}) depicts the temporal evolution of a set of density Fourier modes. For $\beta\simeq$ 2, the model predicts stability of modes satisfying $q \xi>$ 2, consistent with the overall constancy of the black dark line, apart from its small oscillatory character. Unstable modes ($q \xi<$ 2), are also expected to have a finite real part of their complex frequency $\Omega$, meaning that these are not purely growing but rather oscillating, consistent with the red and green lighter lines. Their growth rates, in particular, are shown to be in good agreement with the theoretical model. Furthermore, this regime of exponential growth is observed to last for relatively long times before the instability saturates and nonlinear processes begin to take place. The inset plot of Fig.~(\ref{fig:mode_growth}) depicts exactly the same kind of behavior for a different streaming speed, $\beta\simeq$ 1, where agreement with the theoretical model is also verified.
\par 
While initial stages of the instability are well described by the two-stream model, later stages are essentially dominated by nonlinear effects. We numerically observe these nonlinear stages culminating in a high density of vortices, $n_v$. In the central interaction region, it can be as high as $n_v \sim (q_0 \xi / 2 \pi)^{2} = ( \beta /2 \pi)^2$, equivalent to an average inter-vortex distance of the same order as the spacing between adjacent interference fringes. Faster streamings then result in larger number of vortices. This ability to tune the vortex density may prove to be of interest to the investigation of vortex dynamics. Clustering of large coherent structures of high number of same circulation vortices, for instance, have been shown to form Onsager states~\cite{Simula2014, Groszek2018, Gauthier2019, Johnstone2019}. For the sake of reference, experiments on 2D quantum turbulence in the context of Bose-Einstein condensates typically deal with densities of the order of 10$^{-3}$ vortices per healing length~\cite{Kwon2014, Seo2017}. The mechanism described can, in principle, be used to obtain much higher vortex densities. 
\section{Evidence of Turbulent Cascades}\label{sec:cascades}
\begin{figure*}
\centering
\includegraphics[scale=0.55]{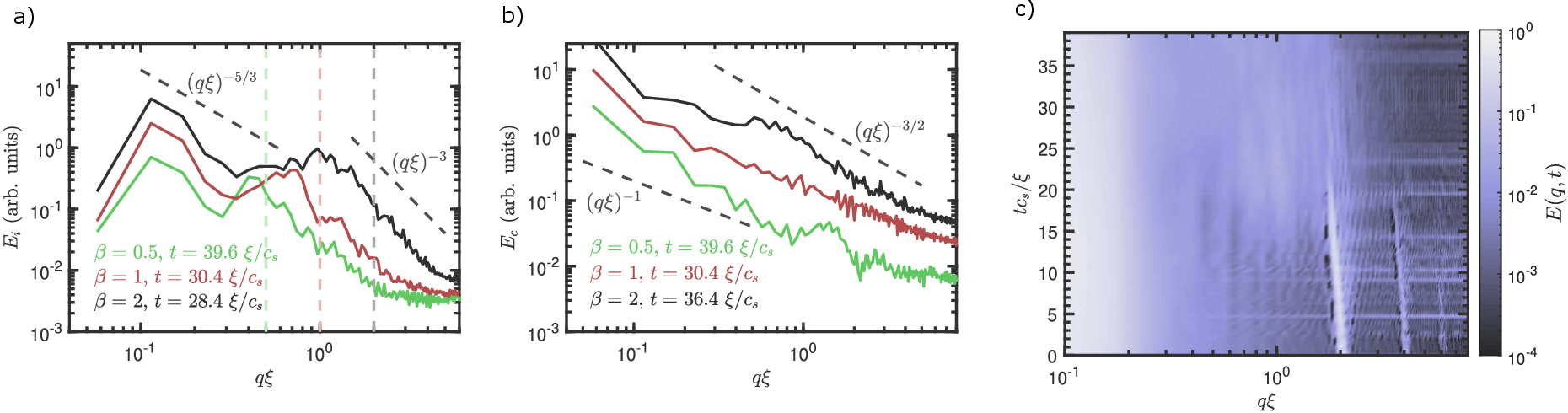}
\caption{(color online) Kinetic energy spectra, decomposed into its incompressible -- panel a) -- and compressible -- panel b) -- components, in the long-time stages of the two-stream instability, for $\beta=$ 0.5, 1 and 2, as before. The spectra have been vertically shifted to allow for a better visualization. The vertical dashed lines indicate the wavenumber associated with the initial interference fringes ($q\xi = \beta$), which can be interpreted as an energy forcing scale, as demonstrated in panel c). Here, the total kinetic energy is plotted as function of both time and wavenumber (for $\beta=2$), clearly demonstrating an initial concentration of energy at ($q\xi = \beta$), which is transferred to both smaller and larger scales as the instability develops.}
\label{fig:turbulent_spectrum}
\end{figure*}
In order to investigate the existence of turbulent cascades during the non-linear stages of the two-stream instability, we decompose the kinetic energy density, defined as $E_k(\mathbf{r}_\perp, t) = (1/2)  \rho (\mathbf{r}_\perp,t ) \lvert \mathbf{v}(\mathbf{r}_\perp,t) \rvert^2$, into its compressible -- $E_k^c (\mathbf{r}_\perp, t)$ -- and incompressible -- $E_k^i (\mathbf{r}_\perp, t)$ -- contributions~\cite{Horng2009, Nore1997, Bradley2012}. To this end, we follow the typical procedure of defining the density weighted velocity field $\mathbf{u} = \sqrt{\rho} \mathbf{v} = \mathbf{u}^c + \mathbf{u}^i$, with $\mathbf{\nabla} \cdot \mathbf{u}^i = 0$ and $\mathbf{\nabla} \times \mathbf{u}^c = 0$, where the explicit time and spacial dependencies were dropped for the sake of notation simplicity. The compressible and incompressible components of the kinetic energy density follow as $E_k^c(\mathbf{r}_\perp,t ) = (1/2) \lvert \mathbf{u}^c(\mathbf{r}_\perp,t ) \rvert^2$ and $E_k^i(\mathbf{r}_\perp,t ) = (1/2) \lvert \mathbf{u}^i(\mathbf{r}_\perp,t ) \rvert^2$, respectively.
\par
The results are depicted in Fig.~(\ref{fig:turbulent_spectrum}). Starting with its incompressible component, the kinetic energy spectrum exhibits a well marked $(q\xi)^{-3}$ decay in the ultraviolet -- $q\xi>1$ -- region. Enstrophy cascades, an hallmark of 2D turbulence, are characterized by a $(q\xi)^{-3}$ decay in the infrared -- $q\xi<1$ -- inertial range~\citep{Bradley2012}. Instead, the decay shown here is a mere consequence of the internal vortex core structure~\cite{Bradley2012, Reeves2012}. We observe a slight shift of this $(q\xi)^{-3}$ structure towards lower wavenumbers as time increases. This is related with the fact that the overall density $\rho$ is slowly decreasing due to the repulsive Kerr nonlinearity, such that the effective healing length is a slowly decreasing function of time. We also find no evidence of a clearly defined Kolmogorov $(q\xi)^{-5/3}$ energy cascade~\cite{Frisch}, which seems to persist in 2D superfluid turbulence under specific conditions~\cite{Bradley2012, Reeves2012}. Contrary to these, the two-stream instability results in a high density of vortex-antivortex pairs, similarly to the so called ultraquantum regime~\cite{Stagg2016}, which does not exhibit Kolmogorov energy cascades. The later are rather associated with regimes of low vortex densities. Energy transport between different scales then result in clusters of same-sign vortices which mimic classical patches of vorticity. This is not observed here, neither the presence of enstrophy cascades, which are also characterized by $(q\xi)^{-3}$ decays, but in the infrared region.
\par
Contrary to its absence in the incompressible component, the compressible part of the kinetic energy spectrum exhibits a $(q\xi)^{-3/2}$ turbulent cascade for all values of $\beta$ investigated here, in the long-time limit. This matches the Kolmogorov-Zakharov weak turbulence spectrum~\cite{Dyachenko1992, Nazarenko2006, Reeves2012}. Such a direct energy cascades is a consequence of non-linear three-wave mixing processes. The $\beta=1$ case seems to slightly deviate from the Kolmogorov-Zakharov scenario. For the sake of completeness, panel b of Fig.~(\ref{fig:turbulent_spectrum}) also depicts what a thermal spectrum would look like, which decays as $(q\xi)^{-1}$. The fitting is still not perfect, and the slight deviations from the $(q\xi)^{-3/2}$ decay may be attributed to anisotropy effects. As an important remark, we reiterate the idea that the initial density modulations resulting from the interference between the two optical fluids effectively act as an energy forcing scale -- see panel c) of Fig.~(\ref{fig:turbulent_spectrum}). We observe a strong initial concentration of kinetic energy at $q \xi = \beta$, which is transferred both to smaller and larger scales as the instability progresses. Thus, the paraxial geometry offers a suitable way of tuning the energy forcing scale by simply setting the relative velocity between the optical fluids. The slight shift of the features in panel c) of Fig.~(\ref{fig:turbulent_spectrum}) to lower wavenumbers is again related with the repulsive character of the system. The regime that follows from the saturation of the two-stream instability is then necessarily transient. In any case, parameters can in principle be chosen such that there is a large gap between the time scales at which the instability develops and energy scales mix, and that at which the mean density varies.  
\par 
As a final remark, in classical turbulence, the dissipation scale sets the high wavenumber (small scale) cutoff of the turbulent cascades. Despite the absence of dissipation in superfluids, an effective viscosity can still be defined upon considering the vortex subsystem, which is allowed to exchange energy with the acoustic bath. Generically however, a lower length scale is usually defined by the vortex core size, $\xi$. Turbulence cascades then typically become relevant in the infra-red region~\cite{Bradley2012}.  
\section{Experimental Considerations}
\par 
Experiments on such paraxial analogues of superfluidity have been conducted in nonlinear crystals~\cite{Shu2012, Sun2012} and atomic vapors driven in the vicinity of a sharp electronic resonance~\cite{Santic2018, Fontaine2018}. The later constitutes a rather interesting platform, where the high level of control over the electronic response allows a precise setting of the model parameters. Recent experiments~\cite{Santic2018} have reported healing lengths of approximately 10 $\mu$m over samples of nearly 1 mm of total spatial extent. This opens the possibility of observing the effects described here. The two-stream instability may be diagnosed by imaging the light exiting the nonlinear medium with a CCD camera, either in real or Fourier space (far-field imaging). The latter provides direct access to the wavenumber of the unstable modes, facilitating the comparison with the theoretical model. The velocity field (phase gradient) can be investigated by interfering the output light with a reference plane wave and analysing the resulting hologram~\cite{Shu2012}. In scenarios of developed turbulence, quantum vortices (phase discontinuities) manifest as fork-like dislocations in the interference pattern~\cite{Lagoudakis2008, Amo2011}. Statistical analysis of the velocity field may require the measurement of the full phase map, which can be retrieved with phase shifting holography~\cite{Sun2012}.
\section{Conclusions}
\par 
We investigated a two-stream instability in paraxial optical fluids. By linearizing a set of hydrodynamic equations, we construct a linear model describing the onset of the instability. The predictions are further corroborated by the numerical integration of the nonlinear Schr\"{o}dinger equation. The latter further describes the nonlinear stages of the instability, where we observe the nucleation of a high density of vortex-antivortex pairs, similar to the so called ultraquantum regime. We investigate the kinetic energy spectrum and discuss the absence of clear evidence for direct energy (Kolmogorov) and enstrophy cascades. The compressible component of the kinetic energy spectrum, however, exhibits a a clear Kolmogorov-Zakharov direct energy turbulent cascade. Interestingly, we demonstrate that the paraxial geometry allows for a straightforward way to selectively excite a particular energy spectral region, thus allowing for the precise engineering of the energy forcing scales.
\section{Acknowledgements}
\par
The present work has received funding from the European Union Horizon 2020 programme, under grant agreement No 820392 (PhoQuS consortium). H. T. thanks Funda\c{c}\~{a}o para a Ci\^{e}ncia e Tecnologia (FCT-Portugal) through Grant No. IF/00433/2015
\bibliographystyle{apsrev4-1}
\bibliography{references}

\begin{thebibliography}{49}%
\makeatletter
\providecommand \@ifxundefined [1]{%
 \@ifx{#1\undefined}
}%
\providecommand \@ifnum [1]{%
 \ifnum #1\expandafter \@firstoftwo
 \else \expandafter \@secondoftwo
 \fi
}%
\providecommand \@ifx [1]{%
 \ifx #1\expandafter \@firstoftwo
 \else \expandafter \@secondoftwo
 \fi
}%
\providecommand \natexlab [1]{#1}%
\providecommand \enquote  [1]{``#1''}%
\providecommand \bibnamefont  [1]{#1}%
\providecommand \bibfnamefont [1]{#1}%
\providecommand \citenamefont [1]{#1}%
\providecommand \href@noop [0]{\@secondoftwo}%
\providecommand \href [0]{\begingroup \@sanitize@url \@href}%
\providecommand \@href[1]{\@@startlink{#1}\@@href}%
\providecommand \@@href[1]{\endgroup#1\@@endlink}%
\providecommand \@sanitize@url [0]{\catcode `\\12\catcode `\$12\catcode
  `\&12\catcode `\#12\catcode `\^12\catcode `\_12\catcode `\%12\relax}%
\providecommand \@@startlink[1]{}%
\providecommand \@@endlink[0]{}%
\providecommand \url  [0]{\begingroup\@sanitize@url \@url }%
\providecommand \@url [1]{\endgroup\@href {#1}{\urlprefix }}%
\providecommand \urlprefix  [0]{URL }%
\providecommand \Eprint [0]{\href }%
\providecommand \doibase [0]{http://dx.doi.org/}%
\providecommand \selectlanguage [0]{\@gobble}%
\providecommand \bibinfo  [0]{\@secondoftwo}%
\providecommand \bibfield  [0]{\@secondoftwo}%
\providecommand \translation [1]{[#1]}%
\providecommand \BibitemOpen [0]{}%
\providecommand \bibitemStop [0]{}%
\providecommand \bibitemNoStop [0]{.\EOS\space}%
\providecommand \EOS [0]{\spacefactor3000\relax}%
\providecommand \BibitemShut  [1]{\csname bibitem#1\endcsname}%
\let\auto@bib@innerbib\@empty
\bibitem [{\citenamefont {Tennekes}\ and\ \citenamefont
  {Lumley}(1972)}]{Tennekes}%
  \BibitemOpen
  \bibfield  {author} {\bibinfo {author} {\bibfnamefont {H.}~\bibnamefont
  {Tennekes}}\ and\ \bibinfo {author} {\bibfnamefont {J.~L.}\ \bibnamefont
  {Lumley}},\ }\href@noop {} {\emph {\bibinfo {title} {A First Course in
  Turbulence}}}\ (\bibinfo  {publisher} {MIT Press},\ \bibinfo {year}
  {1972})\BibitemShut {NoStop}%
\bibitem [{\citenamefont {Pope}(2000)}]{Pope}%
  \BibitemOpen
  \bibfield  {author} {\bibinfo {author} {\bibfnamefont {S.~B.}\ \bibnamefont
  {Pope}},\ }\href@noop {} {\emph {\bibinfo {title} {Turbulent Flows}}}\
  (\bibinfo  {publisher} {Cambridge University Press},\ \bibinfo {year}
  {2000})\BibitemShut {NoStop}%
\bibitem [{\citenamefont {Frisch}(1995)}]{Frisch}%
  \BibitemOpen
  \bibfield  {author} {\bibinfo {author} {\bibfnamefont {U.}~\bibnamefont
  {Frisch}},\ }\href@noop {} {\emph {\bibinfo {title} {Turbulence: The Legacy
  of A.N. Kolmogorov}}}\ (\bibinfo  {publisher} {Cambridge University Press},\
  \bibinfo {year} {1995})\BibitemShut {NoStop}%
\bibitem [{\citenamefont {Falkovich}\ and\ \citenamefont
  {Sreenivasan}(2006)}]{Falkovich2006}%
  \BibitemOpen
  \bibfield  {author} {\bibinfo {author} {\bibfnamefont {G.}~\bibnamefont
  {Falkovich}}\ and\ \bibinfo {author} {\bibfnamefont {K.~R.}\ \bibnamefont
  {Sreenivasan}},\ }\href {\doibase 10.1063/1.2207037} {\bibfield  {journal}
  {\bibinfo  {journal} {Physics Today}\ }\textbf {\bibinfo {volume} {59}},\
  \bibinfo {pages} {43} (\bibinfo {year} {2006})}\BibitemShut {NoStop}%
\bibitem [{\citenamefont {Yepez}\ \emph {et~al.}(2009)\citenamefont {Yepez},
  \citenamefont {Vahala}, \citenamefont {Vahala},\ and\ \citenamefont
  {Soe}}]{Yepez2009}%
  \BibitemOpen
  \bibfield  {author} {\bibinfo {author} {\bibfnamefont {J.}~\bibnamefont
  {Yepez}}, \bibinfo {author} {\bibfnamefont {G.}~\bibnamefont {Vahala}},
  \bibinfo {author} {\bibfnamefont {L.}~\bibnamefont {Vahala}}, \ and\ \bibinfo
  {author} {\bibfnamefont {M.}~\bibnamefont {Soe}},\ }\href {\doibase
  10.1103/PhysRevLett.103.084501} {\bibfield  {journal} {\bibinfo  {journal}
  {Phys. Rev. Lett.}\ }\textbf {\bibinfo {volume} {103}},\ \bibinfo {pages}
  {084501} (\bibinfo {year} {2009})}\BibitemShut {NoStop}%
\bibitem [{\citenamefont {Navon}\ \emph {et~al.}(2016)\citenamefont {Navon},
  \citenamefont {Gaunt}, \citenamefont {Smith},\ and\ \citenamefont
  {Hadzibabic}}]{Navon2016}%
  \BibitemOpen
  \bibfield  {author} {\bibinfo {author} {\bibfnamefont {N.}~\bibnamefont
  {Navon}}, \bibinfo {author} {\bibfnamefont {A.~L.}\ \bibnamefont {Gaunt}},
  \bibinfo {author} {\bibfnamefont {R.~P.}\ \bibnamefont {Smith}}, \ and\
  \bibinfo {author} {\bibfnamefont {Z.}~\bibnamefont {Hadzibabic}},\ }\href
  {http://dx.doi.org/10.1038/nature20114} {\bibfield  {journal} {\bibinfo
  {journal} {Nature}\ }\textbf {\bibinfo {volume} {539}},\ \bibinfo {pages}
  {72} (\bibinfo {year} {2016})},\ \bibinfo {note} {letter}\BibitemShut
  {NoStop}%
\bibitem [{\citenamefont {Onofrio}\ \emph {et~al.}(2000)\citenamefont
  {Onofrio}, \citenamefont {Raman}, \citenamefont {Vogels}, \citenamefont
  {Abo-Shaeer}, \citenamefont {Chikkatur},\ and\ \citenamefont
  {Ketterle}}]{Onofrio2000}%
  \BibitemOpen
  \bibfield  {author} {\bibinfo {author} {\bibfnamefont {R.}~\bibnamefont
  {Onofrio}}, \bibinfo {author} {\bibfnamefont {C.}~\bibnamefont {Raman}},
  \bibinfo {author} {\bibfnamefont {J.~M.}\ \bibnamefont {Vogels}}, \bibinfo
  {author} {\bibfnamefont {J.~R.}\ \bibnamefont {Abo-Shaeer}}, \bibinfo
  {author} {\bibfnamefont {A.~P.}\ \bibnamefont {Chikkatur}}, \ and\ \bibinfo
  {author} {\bibfnamefont {W.}~\bibnamefont {Ketterle}},\ }\href {\doibase
  10.1103/PhysRevLett.85.2228} {\bibfield  {journal} {\bibinfo  {journal}
  {Phys. Rev. Lett.}\ }\textbf {\bibinfo {volume} {85}},\ \bibinfo {pages}
  {2228} (\bibinfo {year} {2000})}\BibitemShut {NoStop}%
\bibitem [{\citenamefont {Desbuquois}\ \emph {et~al.}(2012)\citenamefont
  {Desbuquois}, \citenamefont {Chomaz}, \citenamefont {Yefsah}, \citenamefont
  {Leonard}, \citenamefont {Beugnon}, \citenamefont {Weitenberg},\ and\
  \citenamefont {Dalibard}}]{Desbuquois2012}%
  \BibitemOpen
  \bibfield  {author} {\bibinfo {author} {\bibfnamefont {R.}~\bibnamefont
  {Desbuquois}}, \bibinfo {author} {\bibfnamefont {L.}~\bibnamefont {Chomaz}},
  \bibinfo {author} {\bibfnamefont {T.}~\bibnamefont {Yefsah}}, \bibinfo
  {author} {\bibfnamefont {J.}~\bibnamefont {Leonard}}, \bibinfo {author}
  {\bibfnamefont {J.}~\bibnamefont {Beugnon}}, \bibinfo {author} {\bibfnamefont
  {C.}~\bibnamefont {Weitenberg}}, \ and\ \bibinfo {author} {\bibfnamefont
  {J.}~\bibnamefont {Dalibard}},\ }\href {\doibase 10.1038/nphys2378}
  {\bibfield  {journal} {\bibinfo  {journal} {Nat Phys}\ }\textbf {\bibinfo
  {volume} {8}},\ \bibinfo {pages} {645} (\bibinfo {year} {2012})}\BibitemShut
  {NoStop}%
\bibitem [{\citenamefont {Dalfovo}\ \emph {et~al.}(1999)\citenamefont
  {Dalfovo}, \citenamefont {Giorgini}, \citenamefont {Pitaevskii},\ and\
  \citenamefont {Stringari}}]{Dalfovo1999}%
  \BibitemOpen
  \bibfield  {author} {\bibinfo {author} {\bibfnamefont {F.}~\bibnamefont
  {Dalfovo}}, \bibinfo {author} {\bibfnamefont {S.}~\bibnamefont {Giorgini}},
  \bibinfo {author} {\bibfnamefont {L.~P.}\ \bibnamefont {Pitaevskii}}, \ and\
  \bibinfo {author} {\bibfnamefont {S.}~\bibnamefont {Stringari}},\ }\href
  {\doibase 10.1103/RevModPhys.71.463} {\bibfield  {journal} {\bibinfo
  {journal} {Rev. Mod. Phys.}\ }\textbf {\bibinfo {volume} {71}},\ \bibinfo
  {pages} {463} (\bibinfo {year} {1999})}\BibitemShut {NoStop}%
\bibitem [{\citenamefont {Kasprzak}\ \emph {et~al.}(2006)\citenamefont
  {Kasprzak}, \citenamefont {Richard}, \citenamefont {Kundermann},
  \citenamefont {Baas}, \citenamefont {Jeambrun}, \citenamefont {Keeling},
  \citenamefont {Marchetti}, \citenamefont {Szymanska}, \citenamefont {Andre},
  \citenamefont {Staehli}, \citenamefont {Savona}, \citenamefont {Littlewood},
  \citenamefont {Deveaud},\ and\ \citenamefont {Dang}}]{Kasprzak2006}%
  \BibitemOpen
  \bibfield  {author} {\bibinfo {author} {\bibfnamefont {J.}~\bibnamefont
  {Kasprzak}}, \bibinfo {author} {\bibfnamefont {M.}~\bibnamefont {Richard}},
  \bibinfo {author} {\bibfnamefont {S.}~\bibnamefont {Kundermann}}, \bibinfo
  {author} {\bibfnamefont {A.}~\bibnamefont {Baas}}, \bibinfo {author}
  {\bibfnamefont {P.}~\bibnamefont {Jeambrun}}, \bibinfo {author}
  {\bibfnamefont {J.~M.~J.}\ \bibnamefont {Keeling}}, \bibinfo {author}
  {\bibfnamefont {F.~M.}\ \bibnamefont {Marchetti}}, \bibinfo {author}
  {\bibfnamefont {M.~H.}\ \bibnamefont {Szymanska}}, \bibinfo {author}
  {\bibfnamefont {R.}~\bibnamefont {Andre}}, \bibinfo {author} {\bibfnamefont
  {J.~L.}\ \bibnamefont {Staehli}}, \bibinfo {author} {\bibfnamefont
  {V.}~\bibnamefont {Savona}}, \bibinfo {author} {\bibfnamefont {P.~B.}\
  \bibnamefont {Littlewood}}, \bibinfo {author} {\bibfnamefont
  {B.}~\bibnamefont {Deveaud}}, \ and\ \bibinfo {author} {\bibfnamefont
  {L.~S.}\ \bibnamefont {Dang}},\ }\href {\doibase 10.1038/nature05131}
  {\bibfield  {journal} {\bibinfo  {journal} {Nature}\ }\textbf {\bibinfo
  {volume} {443}},\ \bibinfo {pages} {409} (\bibinfo {year}
  {2006})}\BibitemShut {NoStop}%
\bibitem [{\citenamefont {Deng}\ \emph {et~al.}(2010)\citenamefont {Deng},
  \citenamefont {Haug},\ and\ \citenamefont {Yamamoto}}]{Deng2010}%
  \BibitemOpen
  \bibfield  {author} {\bibinfo {author} {\bibfnamefont {H.}~\bibnamefont
  {Deng}}, \bibinfo {author} {\bibfnamefont {H.}~\bibnamefont {Haug}}, \ and\
  \bibinfo {author} {\bibfnamefont {Y.}~\bibnamefont {Yamamoto}},\ }\href
  {\doibase 10.1103/RevModPhys.82.1489} {\bibfield  {journal} {\bibinfo
  {journal} {Rev. Mod. Phys.}\ }\textbf {\bibinfo {volume} {82}},\ \bibinfo
  {pages} {1489} (\bibinfo {year} {2010})}\BibitemShut {NoStop}%
\bibitem [{\citenamefont {Amo}\ \emph {et~al.}(2009)\citenamefont {Amo},
  \citenamefont {Lefrere}, \citenamefont {Pigeon}, \citenamefont {Adrados},
  \citenamefont {Ciuti}, \citenamefont {Carusotto}, \citenamefont {Houdre},
  \citenamefont {Giacobino},\ and\ \citenamefont {Bramati}}]{Amo2009_1}%
  \BibitemOpen
  \bibfield  {author} {\bibinfo {author} {\bibfnamefont {A.}~\bibnamefont
  {Amo}}, \bibinfo {author} {\bibfnamefont {J.}~\bibnamefont {Lefrere}},
  \bibinfo {author} {\bibfnamefont {S.}~\bibnamefont {Pigeon}}, \bibinfo
  {author} {\bibfnamefont {C.}~\bibnamefont {Adrados}}, \bibinfo {author}
  {\bibfnamefont {C.}~\bibnamefont {Ciuti}}, \bibinfo {author} {\bibfnamefont
  {I.}~\bibnamefont {Carusotto}}, \bibinfo {author} {\bibfnamefont
  {R.}~\bibnamefont {Houdre}}, \bibinfo {author} {\bibfnamefont
  {E.}~\bibnamefont {Giacobino}}, \ and\ \bibinfo {author} {\bibfnamefont
  {A.}~\bibnamefont {Bramati}},\ }\href {\doibase 10.1038/nphys1364} {\bibfield
   {journal} {\bibinfo  {journal} {Nat Phys}\ }\textbf {\bibinfo {volume}
  {5}},\ \bibinfo {pages} {805} (\bibinfo {year} {2009})}\BibitemShut {NoStop}%
\bibitem [{\citenamefont {Lagoudakis}\ \emph {et~al.}(2008)\citenamefont
  {Lagoudakis}, \citenamefont {Wouters}, \citenamefont {Richard}, \citenamefont
  {Baas}, \citenamefont {Carusotto}, \citenamefont {Andre}, \citenamefont
  {Dang},\ and\ \citenamefont {Deveaud-Pledran}}]{Lagoudakis2008}%
  \BibitemOpen
  \bibfield  {author} {\bibinfo {author} {\bibfnamefont {K.~G.}\ \bibnamefont
  {Lagoudakis}}, \bibinfo {author} {\bibfnamefont {M.}~\bibnamefont {Wouters}},
  \bibinfo {author} {\bibfnamefont {M.}~\bibnamefont {Richard}}, \bibinfo
  {author} {\bibfnamefont {A.}~\bibnamefont {Baas}}, \bibinfo {author}
  {\bibfnamefont {I.}~\bibnamefont {Carusotto}}, \bibinfo {author}
  {\bibfnamefont {R.}~\bibnamefont {Andre}}, \bibinfo {author} {\bibfnamefont
  {L.~S.}\ \bibnamefont {Dang}}, \ and\ \bibinfo {author} {\bibfnamefont
  {B.}~\bibnamefont {Deveaud-Pledran}},\ }\href {\doibase 10.1038/nphys1051}
  {\bibfield  {journal} {\bibinfo  {journal} {Nat Phys}\ }\textbf {\bibinfo
  {volume} {4}},\ \bibinfo {pages} {706} (\bibinfo {year} {2008})}\BibitemShut
  {NoStop}%
\bibitem [{\citenamefont {Amo}\ \emph {et~al.}(2011)\citenamefont {Amo},
  \citenamefont {Pigeon}, \citenamefont {Sanvitto}, \citenamefont {Sala},
  \citenamefont {Hivet}, \citenamefont {Carusotto}, \citenamefont {Pisanello},
  \citenamefont {Lem{\'e}nager}, \citenamefont {Houdr{\'e}}, \citenamefont
  {Giacobino}, \citenamefont {Ciuti},\ and\ \citenamefont {Bramati}}]{Amo2011}%
  \BibitemOpen
  \bibfield  {author} {\bibinfo {author} {\bibfnamefont {A.}~\bibnamefont
  {Amo}}, \bibinfo {author} {\bibfnamefont {S.}~\bibnamefont {Pigeon}},
  \bibinfo {author} {\bibfnamefont {D.}~\bibnamefont {Sanvitto}}, \bibinfo
  {author} {\bibfnamefont {V.~G.}\ \bibnamefont {Sala}}, \bibinfo {author}
  {\bibfnamefont {R.}~\bibnamefont {Hivet}}, \bibinfo {author} {\bibfnamefont
  {I.}~\bibnamefont {Carusotto}}, \bibinfo {author} {\bibfnamefont
  {F.}~\bibnamefont {Pisanello}}, \bibinfo {author} {\bibfnamefont
  {G.}~\bibnamefont {Lem{\'e}nager}}, \bibinfo {author} {\bibfnamefont
  {R.}~\bibnamefont {Houdr{\'e}}}, \bibinfo {author} {\bibfnamefont
  {E.}~\bibnamefont {Giacobino}}, \bibinfo {author} {\bibfnamefont
  {C.}~\bibnamefont {Ciuti}}, \ and\ \bibinfo {author} {\bibfnamefont
  {A.}~\bibnamefont {Bramati}},\ }\href {\doibase 10.1126/science.1202307}
  {\bibfield  {journal} {\bibinfo  {journal} {Science}\ }\textbf {\bibinfo
  {volume} {332}},\ \bibinfo {pages} {1167} (\bibinfo {year}
  {2011})}\BibitemShut {NoStop}%
\bibitem [{\citenamefont {Carusotto}\ and\ \citenamefont
  {Ciuti}(2013)}]{Carusotto2013}%
  \BibitemOpen
  \bibfield  {author} {\bibinfo {author} {\bibfnamefont {I.}~\bibnamefont
  {Carusotto}}\ and\ \bibinfo {author} {\bibfnamefont {C.}~\bibnamefont
  {Ciuti}},\ }\href {\doibase 10.1103/RevModPhys.85.299} {\bibfield  {journal}
  {\bibinfo  {journal} {Rev. Mod. Phys.}\ }\textbf {\bibinfo {volume} {85}},\
  \bibinfo {pages} {299} (\bibinfo {year} {2013})}\BibitemShut {NoStop}%
\bibitem [{\citenamefont {Chiao}\ and\ \citenamefont
  {Boyce}(1999)}]{Chiao1999}%
  \BibitemOpen
  \bibfield  {author} {\bibinfo {author} {\bibfnamefont {R.~Y.}\ \bibnamefont
  {Chiao}}\ and\ \bibinfo {author} {\bibfnamefont {J.}~\bibnamefont {Boyce}},\
  }\href {\doibase 10.1103/PhysRevA.60.4114} {\bibfield  {journal} {\bibinfo
  {journal} {Phys. Rev. A}\ }\textbf {\bibinfo {volume} {60}},\ \bibinfo
  {pages} {4114} (\bibinfo {year} {1999})}\BibitemShut {NoStop}%
\bibitem [{\citenamefont {Larr\'e}\ and\ \citenamefont
  {Carusotto}(2015)}]{Larre2015_1}%
  \BibitemOpen
  \bibfield  {author} {\bibinfo {author} {\bibfnamefont {P.-E.}\ \bibnamefont
  {Larr\'e}}\ and\ \bibinfo {author} {\bibfnamefont {I.}~\bibnamefont
  {Carusotto}},\ }\href {\doibase 10.1103/PhysRevA.91.053809} {\bibfield
  {journal} {\bibinfo  {journal} {Phys. Rev. A}\ }\textbf {\bibinfo {volume}
  {91}},\ \bibinfo {pages} {053809} (\bibinfo {year} {2015})}\BibitemShut
  {NoStop}%
\bibitem [{\citenamefont {Vocke}\ \emph {et~al.}(2015)\citenamefont {Vocke},
  \citenamefont {Roger}, \citenamefont {Marino}, \citenamefont {Wright},
  \citenamefont {Carusotto}, \citenamefont {Clerici},\ and\ \citenamefont
  {Faccio}}]{Vocke2015}%
  \BibitemOpen
  \bibfield  {author} {\bibinfo {author} {\bibfnamefont {D.}~\bibnamefont
  {Vocke}}, \bibinfo {author} {\bibfnamefont {T.}~\bibnamefont {Roger}},
  \bibinfo {author} {\bibfnamefont {F.}~\bibnamefont {Marino}}, \bibinfo
  {author} {\bibfnamefont {E.~M.}\ \bibnamefont {Wright}}, \bibinfo {author}
  {\bibfnamefont {I.}~\bibnamefont {Carusotto}}, \bibinfo {author}
  {\bibfnamefont {M.}~\bibnamefont {Clerici}}, \ and\ \bibinfo {author}
  {\bibfnamefont {D.}~\bibnamefont {Faccio}},\ }\href {\doibase
  10.1364/OPTICA.2.000484} {\bibfield  {journal} {\bibinfo  {journal} {Optica}\
  }\textbf {\bibinfo {volume} {2}},\ \bibinfo {pages} {484} (\bibinfo {year}
  {2015})}\BibitemShut {NoStop}%
\bibitem [{\citenamefont {Michel}\ \emph {et~al.}(2018)\citenamefont {Michel},
  \citenamefont {Boughdad}, \citenamefont {Albert}, \citenamefont {Larr{\'e}},\
  and\ \citenamefont {Bellec}}]{Michel2018}%
  \BibitemOpen
  \bibfield  {author} {\bibinfo {author} {\bibfnamefont {C.}~\bibnamefont
  {Michel}}, \bibinfo {author} {\bibfnamefont {O.}~\bibnamefont {Boughdad}},
  \bibinfo {author} {\bibfnamefont {M.}~\bibnamefont {Albert}}, \bibinfo
  {author} {\bibfnamefont {P.-{\'E}.}\ \bibnamefont {Larr{\'e}}}, \ and\
  \bibinfo {author} {\bibfnamefont {M.}~\bibnamefont {Bellec}},\ }\href
  {\doibase 10.1038/s41467-018-04534-9} {\bibfield  {journal} {\bibinfo
  {journal} {Nature Communications}\ }\textbf {\bibinfo {volume} {9}},\
  \bibinfo {pages} {2108} (\bibinfo {year} {2018})}\BibitemShut {NoStop}%
\bibitem [{\citenamefont {Swartzlander}\ and\ \citenamefont
  {Law}(1992)}]{Swartzlander1992}%
  \BibitemOpen
  \bibfield  {author} {\bibinfo {author} {\bibfnamefont {G.~A.}\ \bibnamefont
  {Swartzlander}}\ and\ \bibinfo {author} {\bibfnamefont {C.~T.}\ \bibnamefont
  {Law}},\ }\href {\doibase 10.1103/PhysRevLett.69.2503} {\bibfield  {journal}
  {\bibinfo  {journal} {Phys. Rev. Lett.}\ }\textbf {\bibinfo {volume} {69}},\
  \bibinfo {pages} {2503} (\bibinfo {year} {1992})}\BibitemShut {NoStop}%
\bibitem [{\citenamefont {Jia}\ \emph {et~al.}(2012)\citenamefont {Jia},
  \citenamefont {Haataja},\ and\ \citenamefont {Fleischer}}]{Shu2012}%
  \BibitemOpen
  \bibfield  {author} {\bibinfo {author} {\bibfnamefont {S.}~\bibnamefont
  {Jia}}, \bibinfo {author} {\bibfnamefont {M.}~\bibnamefont {Haataja}}, \ and\
  \bibinfo {author} {\bibfnamefont {J.~W.}\ \bibnamefont {Fleischer}},\ }\href
  {http://stacks.iop.org/1367-2630/14/i=7/a=075009} {\bibfield  {journal}
  {\bibinfo  {journal} {New Journal of Physics}\ }\textbf {\bibinfo {volume}
  {14}},\ \bibinfo {pages} {075009} (\bibinfo {year} {2012})}\BibitemShut
  {NoStop}%
\bibitem [{\citenamefont {Sun}\ \emph {et~al.}(2012)\citenamefont {Sun},
  \citenamefont {Jia}, \citenamefont {Barsi}, \citenamefont {Rica},
  \citenamefont {Picozzi},\ and\ \citenamefont {Fleischer}}]{Sun2012}%
  \BibitemOpen
  \bibfield  {author} {\bibinfo {author} {\bibfnamefont {C.}~\bibnamefont
  {Sun}}, \bibinfo {author} {\bibfnamefont {S.}~\bibnamefont {Jia}}, \bibinfo
  {author} {\bibfnamefont {C.}~\bibnamefont {Barsi}}, \bibinfo {author}
  {\bibfnamefont {S.}~\bibnamefont {Rica}}, \bibinfo {author} {\bibfnamefont
  {A.}~\bibnamefont {Picozzi}}, \ and\ \bibinfo {author} {\bibfnamefont
  {J.~W.}\ \bibnamefont {Fleischer}},\ }\href {\doibase 10.1038/nphys2278}
  {\bibfield  {journal} {\bibinfo  {journal} {Nat Phys}\ }\textbf {\bibinfo
  {volume} {8}},\ \bibinfo {pages} {470} (\bibinfo {year} {2012})}\BibitemShut
  {NoStop}%
\bibitem [{\citenamefont {Bradley}\ and\ \citenamefont
  {Anderson}(2012)}]{Bradley2012}%
  \BibitemOpen
  \bibfield  {author} {\bibinfo {author} {\bibfnamefont {A.~S.}\ \bibnamefont
  {Bradley}}\ and\ \bibinfo {author} {\bibfnamefont {B.~P.}\ \bibnamefont
  {Anderson}},\ }\href {\doibase 10.1103/PhysRevX.2.041001} {\bibfield
  {journal} {\bibinfo  {journal} {Phys. Rev. X}\ }\textbf {\bibinfo {volume}
  {2}},\ \bibinfo {pages} {041001} (\bibinfo {year} {2012})}\BibitemShut
  {NoStop}%
\bibitem [{\citenamefont {Reeves}\ \emph {et~al.}(2013)\citenamefont {Reeves},
  \citenamefont {Billam}, \citenamefont {Anderson},\ and\ \citenamefont
  {Bradley}}]{Reeves2013}%
  \BibitemOpen
  \bibfield  {author} {\bibinfo {author} {\bibfnamefont {M.~T.}\ \bibnamefont
  {Reeves}}, \bibinfo {author} {\bibfnamefont {T.~P.}\ \bibnamefont {Billam}},
  \bibinfo {author} {\bibfnamefont {B.~P.}\ \bibnamefont {Anderson}}, \ and\
  \bibinfo {author} {\bibfnamefont {A.~S.}\ \bibnamefont {Bradley}},\ }\href
  {\doibase 10.1103/PhysRevLett.110.104501} {\bibfield  {journal} {\bibinfo
  {journal} {Phys. Rev. Lett.}\ }\textbf {\bibinfo {volume} {110}},\ \bibinfo
  {pages} {104501} (\bibinfo {year} {2013})}\BibitemShut {NoStop}%
\bibitem [{\citenamefont {Billam}\ \emph {et~al.}(2014)\citenamefont {Billam},
  \citenamefont {Reeves}, \citenamefont {Anderson},\ and\ \citenamefont
  {Bradley}}]{Billam2014}%
  \BibitemOpen
  \bibfield  {author} {\bibinfo {author} {\bibfnamefont {T.~P.}\ \bibnamefont
  {Billam}}, \bibinfo {author} {\bibfnamefont {M.~T.}\ \bibnamefont {Reeves}},
  \bibinfo {author} {\bibfnamefont {B.~P.}\ \bibnamefont {Anderson}}, \ and\
  \bibinfo {author} {\bibfnamefont {A.~S.}\ \bibnamefont {Bradley}},\ }\href
  {\doibase 10.1103/PhysRevLett.112.145301} {\bibfield  {journal} {\bibinfo
  {journal} {Phys. Rev. Lett.}\ }\textbf {\bibinfo {volume} {112}},\ \bibinfo
  {pages} {145301} (\bibinfo {year} {2014})}\BibitemShut {NoStop}%
\bibitem [{\citenamefont {Fontaine}\ \emph {et~al.}(2018)\citenamefont
  {Fontaine}, \citenamefont {Bienaim\'e}, \citenamefont {Pigeon}, \citenamefont
  {Giacobino}, \citenamefont {Bramati},\ and\ \citenamefont
  {Glorieux}}]{Fontaine2018}%
  \BibitemOpen
  \bibfield  {author} {\bibinfo {author} {\bibfnamefont {Q.}~\bibnamefont
  {Fontaine}}, \bibinfo {author} {\bibfnamefont {T.}~\bibnamefont
  {Bienaim\'e}}, \bibinfo {author} {\bibfnamefont {S.}~\bibnamefont {Pigeon}},
  \bibinfo {author} {\bibfnamefont {E.}~\bibnamefont {Giacobino}}, \bibinfo
  {author} {\bibfnamefont {A.}~\bibnamefont {Bramati}}, \ and\ \bibinfo
  {author} {\bibfnamefont {Q.}~\bibnamefont {Glorieux}},\ }\href {\doibase
  10.1103/PhysRevLett.121.183604} {\bibfield  {journal} {\bibinfo  {journal}
  {Phys. Rev. Lett.}\ }\textbf {\bibinfo {volume} {121}},\ \bibinfo {pages}
  {183604} (\bibinfo {year} {2018})}\BibitemShut {NoStop}%
\bibitem [{\citenamefont {\ifmmode \check{S}\else
  \v{S}\fi{}anti\ifmmode~\acute{c}\else \'{c}\fi{}}\ \emph
  {et~al.}(2018)\citenamefont {\ifmmode \check{S}\else
  \v{S}\fi{}anti\ifmmode~\acute{c}\else \'{c}\fi{}}, \citenamefont {Fusaro},
  \citenamefont {Salem}, \citenamefont {Garnier}, \citenamefont {Picozzi},\
  and\ \citenamefont {Kaiser}}]{Santic2018}%
  \BibitemOpen
  \bibfield  {author} {\bibinfo {author} {\bibfnamefont {N.}~\bibnamefont
  {\ifmmode \check{S}\else \v{S}\fi{}anti\ifmmode~\acute{c}\else \'{c}\fi{}}},
  \bibinfo {author} {\bibfnamefont {A.}~\bibnamefont {Fusaro}}, \bibinfo
  {author} {\bibfnamefont {S.}~\bibnamefont {Salem}}, \bibinfo {author}
  {\bibfnamefont {J.}~\bibnamefont {Garnier}}, \bibinfo {author} {\bibfnamefont
  {A.}~\bibnamefont {Picozzi}}, \ and\ \bibinfo {author} {\bibfnamefont
  {R.}~\bibnamefont {Kaiser}},\ }\href {\doibase
  10.1103/PhysRevLett.120.055301} {\bibfield  {journal} {\bibinfo  {journal}
  {Phys. Rev. Lett.}\ }\textbf {\bibinfo {volume} {120}},\ \bibinfo {pages}
  {055301} (\bibinfo {year} {2018})}\BibitemShut {NoStop}%
\bibitem [{\citenamefont {Hall}\ \emph {et~al.}(1998)\citenamefont {Hall},
  \citenamefont {Matthews}, \citenamefont {Ensher}, \citenamefont {Wieman},\
  and\ \citenamefont {Cornell}}]{Hall1998}%
  \BibitemOpen
  \bibfield  {author} {\bibinfo {author} {\bibfnamefont {D.~S.}\ \bibnamefont
  {Hall}}, \bibinfo {author} {\bibfnamefont {M.~R.}\ \bibnamefont {Matthews}},
  \bibinfo {author} {\bibfnamefont {J.~R.}\ \bibnamefont {Ensher}}, \bibinfo
  {author} {\bibfnamefont {C.~E.}\ \bibnamefont {Wieman}}, \ and\ \bibinfo
  {author} {\bibfnamefont {E.~A.}\ \bibnamefont {Cornell}},\ }\href {\doibase
  10.1103/PhysRevLett.81.1539} {\bibfield  {journal} {\bibinfo  {journal}
  {Phys. Rev. Lett.}\ }\textbf {\bibinfo {volume} {81}},\ \bibinfo {pages}
  {1539} (\bibinfo {year} {1998})}\BibitemShut {NoStop}%
\bibitem [{\citenamefont {Lin}\ \emph {et~al.}(2011)\citenamefont {Lin},
  \citenamefont {Jim{\'e}nez-Garc{\'i}a},\ and\ \citenamefont
  {Spielman}}]{Lin2011}%
  \BibitemOpen
  \bibfield  {author} {\bibinfo {author} {\bibfnamefont {Y.-J.}\ \bibnamefont
  {Lin}}, \bibinfo {author} {\bibfnamefont {K.}~\bibnamefont
  {Jim{\'e}nez-Garc{\'i}a}}, \ and\ \bibinfo {author} {\bibfnamefont {I.~B.}\
  \bibnamefont {Spielman}},\ }\href {\doibase 10.1038/nature09887} {\bibfield
  {journal} {\bibinfo  {journal} {Nature}\ }\textbf {\bibinfo {volume} {471}},\
  \bibinfo {pages} {83} (\bibinfo {year} {2011})}\BibitemShut {NoStop}%
\bibitem [{\citenamefont {Abad}\ \emph {et~al.}(2015)\citenamefont {Abad},
  \citenamefont {Recati}, \citenamefont {Stringari},\ and\ \citenamefont
  {Chevy}}]{Abad2015}%
  \BibitemOpen
  \bibfield  {author} {\bibinfo {author} {\bibfnamefont {M.}~\bibnamefont
  {Abad}}, \bibinfo {author} {\bibfnamefont {A.}~\bibnamefont {Recati}},
  \bibinfo {author} {\bibfnamefont {S.}~\bibnamefont {Stringari}}, \ and\
  \bibinfo {author} {\bibfnamefont {F.}~\bibnamefont {Chevy}},\ }\href
  {\doibase 10.1140/epjd/e2015-50851-y} {\bibfield  {journal} {\bibinfo
  {journal} {The European Physical Journal D}\ }\textbf {\bibinfo {volume}
  {69}},\ \bibinfo {pages} {126} (\bibinfo {year} {2015})}\BibitemShut
  {NoStop}%
\bibitem [{\citenamefont {Nicholson}(1983)}]{Nicholson}%
  \BibitemOpen
  \bibfield  {author} {\bibinfo {author} {\bibfnamefont {D.~R.}\ \bibnamefont
  {Nicholson}},\ }\href@noop {} {\emph {\bibinfo {title} {Introduction to
  Plasma Theory}}}\ (\bibinfo  {publisher} {Wiley, New York},\ \bibinfo {year}
  {1983})\BibitemShut {NoStop}%
\bibitem [{\citenamefont {Mendon\c{c}a}\ \emph {et~al.}(2008)\citenamefont
  {Mendon\c{c}a}, \citenamefont {Serbeto},\ and\ \citenamefont
  {Shukla}}]{Mendonca2008}%
  \BibitemOpen
  \bibfield  {author} {\bibinfo {author} {\bibfnamefont {J.}~\bibnamefont
  {Mendon\c{c}a}}, \bibinfo {author} {\bibfnamefont {A.}~\bibnamefont
  {Serbeto}}, \ and\ \bibinfo {author} {\bibfnamefont {P.}~\bibnamefont
  {Shukla}},\ }\href {\doibase
  http://dx.doi.org/10.1016/j.physleta.2007.11.021} {\bibfield  {journal}
  {\bibinfo  {journal} {Physics Letters A}\ }\textbf {\bibinfo {volume}
  {372}},\ \bibinfo {pages} {2311 } (\bibinfo {year} {2008})}\BibitemShut
  {NoStop}%
\bibitem [{\citenamefont {Ter\ifmmode~\mbox{\c{c}}\else \c{c}\fi{}as}\ \emph
  {et~al.}(2009)\citenamefont {Ter\ifmmode~\mbox{\c{c}}\else \c{c}\fi{}as},
  \citenamefont {Mendon\ifmmode~\mbox{\c{c}}\else \c{c}\fi{}a},\ and\
  \citenamefont {Robb}}]{Tercas2009}%
  \BibitemOpen
  \bibfield  {author} {\bibinfo {author} {\bibfnamefont {H.}~\bibnamefont
  {Ter\ifmmode~\mbox{\c{c}}\else \c{c}\fi{}as}}, \bibinfo {author}
  {\bibfnamefont {J.~T.}\ \bibnamefont {Mendon\ifmmode~\mbox{\c{c}}\else
  \c{c}\fi{}a}}, \ and\ \bibinfo {author} {\bibfnamefont {G.~R.~M.}\
  \bibnamefont {Robb}},\ }\href {\doibase 10.1103/PhysRevA.79.065601}
  {\bibfield  {journal} {\bibinfo  {journal} {Phys. Rev. A}\ }\textbf {\bibinfo
  {volume} {79}},\ \bibinfo {pages} {065601} (\bibinfo {year}
  {2009})}\BibitemShut {NoStop}%
\bibitem [{\citenamefont {Guo}\ \emph {et~al.}(2010)\citenamefont {Guo},
  \citenamefont {Cahn}, \citenamefont {Nikkel}, \citenamefont {Vinen},\ and\
  \citenamefont {McKinsey}}]{Guo2010}%
  \BibitemOpen
  \bibfield  {author} {\bibinfo {author} {\bibfnamefont {W.}~\bibnamefont
  {Guo}}, \bibinfo {author} {\bibfnamefont {S.~B.}\ \bibnamefont {Cahn}},
  \bibinfo {author} {\bibfnamefont {J.~A.}\ \bibnamefont {Nikkel}}, \bibinfo
  {author} {\bibfnamefont {W.~F.}\ \bibnamefont {Vinen}}, \ and\ \bibinfo
  {author} {\bibfnamefont {D.~N.}\ \bibnamefont {McKinsey}},\ }\href {\doibase
  10.1103/PhysRevLett.105.045301} {\bibfield  {journal} {\bibinfo  {journal}
  {Phys. Rev. Lett.}\ }\textbf {\bibinfo {volume} {105}},\ \bibinfo {pages}
  {045301} (\bibinfo {year} {2010})}\BibitemShut {NoStop}%
\bibitem [{\citenamefont {Lushnikov}\ and\ \citenamefont
  {Zubarev}(2018)}]{Lushnikov2018}%
  \BibitemOpen
  \bibfield  {author} {\bibinfo {author} {\bibfnamefont {P.~M.}\ \bibnamefont
  {Lushnikov}}\ and\ \bibinfo {author} {\bibfnamefont {N.~M.}\ \bibnamefont
  {Zubarev}},\ }\href {\doibase 10.1103/PhysRevLett.120.204504} {\bibfield
  {journal} {\bibinfo  {journal} {Phys. Rev. Lett.}\ }\textbf {\bibinfo
  {volume} {120}},\ \bibinfo {pages} {204504} (\bibinfo {year}
  {2018})}\BibitemShut {NoStop}%
\bibitem [{\citenamefont {Ishino}\ \emph {et~al.}(2011)\citenamefont {Ishino},
  \citenamefont {Takeuchi},\ and\ \citenamefont {Tsubota}}]{Ishino2011}%
  \BibitemOpen
  \bibfield  {author} {\bibinfo {author} {\bibfnamefont {S.}~\bibnamefont
  {Ishino}}, \bibinfo {author} {\bibfnamefont {H.}~\bibnamefont {Takeuchi}}, \
  and\ \bibinfo {author} {\bibfnamefont {M.}~\bibnamefont {Tsubota}},\ }\href
  {\doibase 10.1007/s10909-010-0324-y} {\bibfield  {journal} {\bibinfo
  {journal} {Journal of Low Temperature Physics}\ }\textbf {\bibinfo {volume}
  {162}},\ \bibinfo {pages} {361} (\bibinfo {year} {2011})}\BibitemShut
  {NoStop}%
\bibitem [{\citenamefont {Antoine}\ \emph {et~al.}(2013)\citenamefont
  {Antoine}, \citenamefont {Bao},\ and\ \citenamefont {Besse}}]{Antoine2013}%
  \BibitemOpen
  \bibfield  {author} {\bibinfo {author} {\bibfnamefont {X.}~\bibnamefont
  {Antoine}}, \bibinfo {author} {\bibfnamefont {W.}~\bibnamefont {Bao}}, \ and\
  \bibinfo {author} {\bibfnamefont {C.}~\bibnamefont {Besse}},\ }\href
  {\doibase https://doi.org/10.1016/j.cpc.2013.07.012} {\bibfield  {journal}
  {\bibinfo  {journal} {Computer Physics Communications}\ }\textbf {\bibinfo
  {volume} {184}},\ \bibinfo {pages} {2621 } (\bibinfo {year}
  {2013})}\BibitemShut {NoStop}%
\bibitem [{\citenamefont {Simula}\ \emph {et~al.}(2014)\citenamefont {Simula},
  \citenamefont {Davis},\ and\ \citenamefont {Helmerson}}]{Simula2014}%
  \BibitemOpen
  \bibfield  {author} {\bibinfo {author} {\bibfnamefont {T.}~\bibnamefont
  {Simula}}, \bibinfo {author} {\bibfnamefont {M.~J.}\ \bibnamefont {Davis}}, \
  and\ \bibinfo {author} {\bibfnamefont {K.}~\bibnamefont {Helmerson}},\ }\href
  {\doibase 10.1103/PhysRevLett.113.165302} {\bibfield  {journal} {\bibinfo
  {journal} {Phys. Rev. Lett.}\ }\textbf {\bibinfo {volume} {113}},\ \bibinfo
  {pages} {165302} (\bibinfo {year} {2014})}\BibitemShut {NoStop}%
\bibitem [{\citenamefont {Groszek}\ \emph {et~al.}(2018)\citenamefont
  {Groszek}, \citenamefont {Davis}, \citenamefont {Paganin}, \citenamefont
  {Helmerson},\ and\ \citenamefont {Simula}}]{Groszek2018}%
  \BibitemOpen
  \bibfield  {author} {\bibinfo {author} {\bibfnamefont {A.~J.}\ \bibnamefont
  {Groszek}}, \bibinfo {author} {\bibfnamefont {M.~J.}\ \bibnamefont {Davis}},
  \bibinfo {author} {\bibfnamefont {D.~M.}\ \bibnamefont {Paganin}}, \bibinfo
  {author} {\bibfnamefont {K.}~\bibnamefont {Helmerson}}, \ and\ \bibinfo
  {author} {\bibfnamefont {T.~P.}\ \bibnamefont {Simula}},\ }\href {\doibase
  10.1103/PhysRevLett.120.034504} {\bibfield  {journal} {\bibinfo  {journal}
  {Phys. Rev. Lett.}\ }\textbf {\bibinfo {volume} {120}},\ \bibinfo {pages}
  {034504} (\bibinfo {year} {2018})}\BibitemShut {NoStop}%
\bibitem [{\citenamefont {Gauthier}\ \emph {et~al.}(2019)\citenamefont
  {Gauthier}, \citenamefont {Reeves}, \citenamefont {Yu}, \citenamefont
  {Bradley}, \citenamefont {Baker}, \citenamefont {Bell}, \citenamefont
  {Rubinsztein-Dunlop}, \citenamefont {Davis},\ and\ \citenamefont
  {Neely}}]{Gauthier2019}%
  \BibitemOpen
  \bibfield  {author} {\bibinfo {author} {\bibfnamefont {G.}~\bibnamefont
  {Gauthier}}, \bibinfo {author} {\bibfnamefont {M.~T.}\ \bibnamefont
  {Reeves}}, \bibinfo {author} {\bibfnamefont {X.}~\bibnamefont {Yu}}, \bibinfo
  {author} {\bibfnamefont {A.~S.}\ \bibnamefont {Bradley}}, \bibinfo {author}
  {\bibfnamefont {M.~A.}\ \bibnamefont {Baker}}, \bibinfo {author}
  {\bibfnamefont {T.~A.}\ \bibnamefont {Bell}}, \bibinfo {author}
  {\bibfnamefont {H.}~\bibnamefont {Rubinsztein-Dunlop}}, \bibinfo {author}
  {\bibfnamefont {M.~J.}\ \bibnamefont {Davis}}, \ and\ \bibinfo {author}
  {\bibfnamefont {T.~W.}\ \bibnamefont {Neely}},\ }\href {\doibase
  10.1126/science.aat5718} {\bibfield  {journal} {\bibinfo  {journal}
  {Science}\ }\textbf {\bibinfo {volume} {364}},\ \bibinfo {pages} {1264}
  (\bibinfo {year} {2019})}\BibitemShut {NoStop}%
\bibitem [{\citenamefont {Johnstone}\ \emph {et~al.}(2019)\citenamefont
  {Johnstone}, \citenamefont {Groszek}, \citenamefont {Starkey}, \citenamefont
  {Billington}, \citenamefont {Simula},\ and\ \citenamefont
  {Helmerson}}]{Johnstone2019}%
  \BibitemOpen
  \bibfield  {author} {\bibinfo {author} {\bibfnamefont {S.~P.}\ \bibnamefont
  {Johnstone}}, \bibinfo {author} {\bibfnamefont {A.~J.}\ \bibnamefont
  {Groszek}}, \bibinfo {author} {\bibfnamefont {P.~T.}\ \bibnamefont
  {Starkey}}, \bibinfo {author} {\bibfnamefont {C.~J.}\ \bibnamefont
  {Billington}}, \bibinfo {author} {\bibfnamefont {T.~P.}\ \bibnamefont
  {Simula}}, \ and\ \bibinfo {author} {\bibfnamefont {K.}~\bibnamefont
  {Helmerson}},\ }\href {\doibase 10.1126/science.aat5793} {\bibfield
  {journal} {\bibinfo  {journal} {Science}\ }\textbf {\bibinfo {volume}
  {364}},\ \bibinfo {pages} {1267} (\bibinfo {year} {2019})}\BibitemShut
  {NoStop}%
\bibitem [{\citenamefont {Kwon}\ \emph {et~al.}(2014)\citenamefont {Kwon},
  \citenamefont {Moon}, \citenamefont {Choi}, \citenamefont {Seo},\ and\
  \citenamefont {Shin}}]{Kwon2014}%
  \BibitemOpen
  \bibfield  {author} {\bibinfo {author} {\bibfnamefont {W.~J.}\ \bibnamefont
  {Kwon}}, \bibinfo {author} {\bibfnamefont {G.}~\bibnamefont {Moon}}, \bibinfo
  {author} {\bibfnamefont {J.-y.}\ \bibnamefont {Choi}}, \bibinfo {author}
  {\bibfnamefont {S.~W.}\ \bibnamefont {Seo}}, \ and\ \bibinfo {author}
  {\bibfnamefont {Y.-i.}\ \bibnamefont {Shin}},\ }\href {\doibase
  10.1103/PhysRevA.90.063627} {\bibfield  {journal} {\bibinfo  {journal} {Phys.
  Rev. A}\ }\textbf {\bibinfo {volume} {90}},\ \bibinfo {pages} {063627}
  (\bibinfo {year} {2014})}\BibitemShut {NoStop}%
\bibitem [{\citenamefont {Seo}\ \emph {et~al.}(2017)\citenamefont {Seo},
  \citenamefont {Ko}, \citenamefont {Kim},\ and\ \citenamefont
  {Shin}}]{Seo2017}%
  \BibitemOpen
  \bibfield  {author} {\bibinfo {author} {\bibfnamefont {S.~W.}\ \bibnamefont
  {Seo}}, \bibinfo {author} {\bibfnamefont {B.}~\bibnamefont {Ko}}, \bibinfo
  {author} {\bibfnamefont {J.~H.}\ \bibnamefont {Kim}}, \ and\ \bibinfo
  {author} {\bibfnamefont {Y.}~\bibnamefont {Shin}},\ }\href {\doibase
  10.1038/s41598-017-04122-9} {\bibfield  {journal} {\bibinfo  {journal}
  {Scientific Reports}\ }\textbf {\bibinfo {volume} {7}},\ \bibinfo {pages}
  {4587} (\bibinfo {year} {2017})}\BibitemShut {NoStop}%
\bibitem [{\citenamefont {Horng}\ \emph {et~al.}(2009)\citenamefont {Horng},
  \citenamefont {Hsueh}, \citenamefont {Su}, \citenamefont {Kao},\ and\
  \citenamefont {Gou}}]{Horng2009}%
  \BibitemOpen
  \bibfield  {author} {\bibinfo {author} {\bibfnamefont {T.-L.}\ \bibnamefont
  {Horng}}, \bibinfo {author} {\bibfnamefont {C.-H.}\ \bibnamefont {Hsueh}},
  \bibinfo {author} {\bibfnamefont {S.-W.}\ \bibnamefont {Su}}, \bibinfo
  {author} {\bibfnamefont {Y.-M.}\ \bibnamefont {Kao}}, \ and\ \bibinfo
  {author} {\bibfnamefont {S.-C.}\ \bibnamefont {Gou}},\ }\href {\doibase
  10.1103/PhysRevA.80.023618} {\bibfield  {journal} {\bibinfo  {journal} {Phys.
  Rev. A}\ }\textbf {\bibinfo {volume} {80}},\ \bibinfo {pages} {023618}
  (\bibinfo {year} {2009})}\BibitemShut {NoStop}%
\bibitem [{\citenamefont {Nore}\ \emph {et~al.}(1997)\citenamefont {Nore},
  \citenamefont {Abid},\ and\ \citenamefont {Brachet}}]{Nore1997}%
  \BibitemOpen
  \bibfield  {author} {\bibinfo {author} {\bibfnamefont {C.}~\bibnamefont
  {Nore}}, \bibinfo {author} {\bibfnamefont {M.}~\bibnamefont {Abid}}, \ and\
  \bibinfo {author} {\bibfnamefont {M.~E.}\ \bibnamefont {Brachet}},\ }\href
  {\doibase 10.1103/PhysRevLett.78.3896} {\bibfield  {journal} {\bibinfo
  {journal} {Phys. Rev. Lett.}\ }\textbf {\bibinfo {volume} {78}},\ \bibinfo
  {pages} {3896} (\bibinfo {year} {1997})}\BibitemShut {NoStop}%
\bibitem [{\citenamefont {Reeves}\ \emph {et~al.}(2012)\citenamefont {Reeves},
  \citenamefont {Anderson},\ and\ \citenamefont {Bradley}}]{Reeves2012}%
  \BibitemOpen
  \bibfield  {author} {\bibinfo {author} {\bibfnamefont {M.~T.}\ \bibnamefont
  {Reeves}}, \bibinfo {author} {\bibfnamefont {B.~P.}\ \bibnamefont
  {Anderson}}, \ and\ \bibinfo {author} {\bibfnamefont {A.~S.}\ \bibnamefont
  {Bradley}},\ }\href {\doibase 10.1103/PhysRevA.86.053621} {\bibfield
  {journal} {\bibinfo  {journal} {Phys. Rev. A}\ }\textbf {\bibinfo {volume}
  {86}},\ \bibinfo {pages} {053621} (\bibinfo {year} {2012})}\BibitemShut
  {NoStop}%
\bibitem [{\citenamefont {Stagg}\ \emph {et~al.}(2016)\citenamefont {Stagg},
  \citenamefont {Parker},\ and\ \citenamefont {Barenghi}}]{Stagg2016}%
  \BibitemOpen
  \bibfield  {author} {\bibinfo {author} {\bibfnamefont {G.~W.}\ \bibnamefont
  {Stagg}}, \bibinfo {author} {\bibfnamefont {N.~G.}\ \bibnamefont {Parker}}, \
  and\ \bibinfo {author} {\bibfnamefont {C.~F.}\ \bibnamefont {Barenghi}},\
  }\href {\doibase 10.1103/PhysRevA.94.053632} {\bibfield  {journal} {\bibinfo
  {journal} {Phys. Rev. A}\ }\textbf {\bibinfo {volume} {94}},\ \bibinfo
  {pages} {053632} (\bibinfo {year} {2016})}\BibitemShut {NoStop}%
\bibitem [{\citenamefont {Dyachenko}\ \emph {et~al.}(1992)\citenamefont
  {Dyachenko}, \citenamefont {Newell}, \citenamefont {Pushkarev},\ and\
  \citenamefont {Zakharov}}]{Dyachenko1992}%
  \BibitemOpen
  \bibfield  {author} {\bibinfo {author} {\bibfnamefont {S.}~\bibnamefont
  {Dyachenko}}, \bibinfo {author} {\bibfnamefont {A.}~\bibnamefont {Newell}},
  \bibinfo {author} {\bibfnamefont {A.}~\bibnamefont {Pushkarev}}, \ and\
  \bibinfo {author} {\bibfnamefont {V.}~\bibnamefont {Zakharov}},\ }\href
  {\doibase https://doi.org/10.1016/0167-2789(92)90090-A} {\bibfield  {journal}
  {\bibinfo  {journal} {Physica D: Nonlinear Phenomena}\ }\textbf {\bibinfo
  {volume} {57}},\ \bibinfo {pages} {96 } (\bibinfo {year} {1992})}\BibitemShut
  {NoStop}%
\bibitem [{\citenamefont {Nazarenko}\ and\ \citenamefont
  {Onorato}(2006)}]{Nazarenko2006}%
  \BibitemOpen
  \bibfield  {author} {\bibinfo {author} {\bibfnamefont {S.}~\bibnamefont
  {Nazarenko}}\ and\ \bibinfo {author} {\bibfnamefont {M.}~\bibnamefont
  {Onorato}},\ }\href {\doibase https://doi.org/10.1016/j.physd.2006.05.007}
  {\bibfield  {journal} {\bibinfo  {journal} {Physica D: Nonlinear Phenomena}\
  }\textbf {\bibinfo {volume} {219}},\ \bibinfo {pages} {1 } (\bibinfo {year}
  {2006})}\BibitemShut {NoStop}%
\end{thebibliography}%
\bibliographystyle{apsrev4-1}
\end{document}